\documentclass[a4paper, amsfonts, amssymb, amsmath, reprint, showkeys, nofootinbib]{revtex4-2}
\usepackage[english]{babel}
\usepackage[utf8]{inputenc}
\usepackage[colorinlistoftodos, color=green!40, prependcaption]{todonotes}
\usepackage{amsthm}
\usepackage{mathtools}
\usepackage{physics}
\usepackage{xcolor}
\usepackage{graphicx}
\usepackage[left=23mm,right=13mm,top=35mm,columnsep=15pt]{geometry} 
\usepackage{adjustbox}
\usepackage{placeins}
\usepackage[T1]{fontenc}
\usepackage{lipsum}
\usepackage{csquotes}
\usepackage[pdftex, pdftitle={Article}, pdfauthor={Author}]{hyperref} 
\begin{document}
\title{Self-calibrating gas pressure sensor with a 10-decade measurement range}

\author{\textsuperscript{1}Christoph Reinhardt}
    \thanks{C.R. and H.M. contributed equally to this work. \\ 
    To whom correspondence may be addressed:\\
    \href{mailto:christoph.reinhardt@desy.de}{christoph.reinhardt@desy.de} \\
    \href{mailto:hossein.masalehdan@physnet.uni-hamburg.de}{hossein.masalehdan@physnet.uni-hamburg.de}}
    \affiliation{\textsuperscript{1}Deutsches Elektronen-Synchrotron DESY, Notkestr. 85, 22607 Hamburg, Germany \\ \textsuperscript{2}Institut für Quantenphysik (IQP) \& Zentrum für Optische Quantentechnologien (ZOQ), Universität Hamburg, 22761 Hamburg, Germany}
\author{\textsuperscript{2}Hossein Masalehdan} 
    \thanks{C.R. and H.M. contributed equally to this work. \\ 
    To whom correspondence may be addressed:\\
    \href{mailto:christoph.reinhardt@desy.de}{christoph.reinhardt@desy.de} \\
    \href{mailto:hossein.masalehdan@physnet.uni-hamburg.de}{hossein.masalehdan@physnet.uni-hamburg.de}}
    \affiliation{\textsuperscript{1}Deutsches Elektronen-Synchrotron DESY, Notkestr. 85, 22607 Hamburg, Germany \\ \textsuperscript{2}Institut für Quantenphysik (IQP) \& Zentrum für Optische Quantentechnologien (ZOQ), Universität Hamburg, 22761 Hamburg, Germany}
\author{\textsuperscript{1}Sandy Croatto}
    \affiliation{\textsuperscript{1}Deutsches Elektronen-Synchrotron DESY, Notkestr. 85, 22607 Hamburg, Germany \\ \textsuperscript{2}Institut für Quantenphysik (IQP) \& Zentrum für Optische Quantentechnologien (ZOQ), Universität Hamburg, 22761 Hamburg, Germany}
\author{\textsuperscript{2}Alexander Franke} 
    \affiliation{\textsuperscript{1}Deutsches Elektronen-Synchrotron DESY, Notkestr. 85, 22607 Hamburg, Germany \\ \textsuperscript{2}Institut für Quantenphysik (IQP) \& Zentrum für Optische Quantentechnologien (ZOQ), Universität Hamburg, 22761 Hamburg, Germany}
\author{\textsuperscript{1}Moritz B. K. Kunze}
    \affiliation{\textsuperscript{1}Deutsches Elektronen-Synchrotron DESY, Notkestr. 85, 22607 Hamburg, Germany \\ \textsuperscript{2}Institut für Quantenphysik (IQP) \& Zentrum für Optische Quantentechnologien (ZOQ), Universität Hamburg, 22761 Hamburg, Germany}
\author{\textsuperscript{1}Jörn Schaffran}
    \affiliation{\textsuperscript{1}Deutsches Elektronen-Synchrotron DESY, Notkestr. 85, 22607 Hamburg, Germany \\ \textsuperscript{2}Institut für Quantenphysik (IQP) \& Zentrum für Optische Quantentechnologien (ZOQ), Universität Hamburg, 22761 Hamburg, Germany}
\author{\textsuperscript{2}Nils Sültmann} 
    \affiliation{\textsuperscript{1}Deutsches Elektronen-Synchrotron DESY, Notkestr. 85, 22607 Hamburg, Germany \\ \textsuperscript{2}Institut für Quantenphysik (IQP) \& Zentrum für Optische Quantentechnologien (ZOQ), Universität Hamburg, 22761 Hamburg, Germany}
\author{\textsuperscript{1}Axel Lindner}
    \affiliation{\textsuperscript{1}Deutsches Elektronen-Synchrotron DESY, Notkestr. 85, 22607 Hamburg, Germany \\ \textsuperscript{2}Institut für Quantenphysik (IQP) \& Zentrum für Optische Quantentechnologien (ZOQ), Universität Hamburg, 22761 Hamburg, Germany}
\author{\textsuperscript{2}Roman Schnabel} 
    \affiliation{\textsuperscript{1}Deutsches Elektronen-Synchrotron DESY, Notkestr. 85, 22607 Hamburg, Germany \\ \textsuperscript{2}Institut für Quantenphysik (IQP) \& Zentrum für Optische Quantentechnologien (ZOQ), Universität Hamburg, 22761 Hamburg, Germany}

\date{\today} 

\begin{abstract}
Recent years have seen a rapid reduction in the intrinsic loss of nanomechanical resonators (i.e., chip-scale mechanical oscillators).
As a result, these devices become increasingly sensitive to the friction exerted by smallest amounts of gas.
Here, we present the pressure-dependency of a nanomechanical trampoline resonator's quality factor $Q$ over ten decades, from $10^{-7}$ to $10^{3}\,\mathrm{mbar}$.
We find that the measured behavior is well-described by a model combining analytical and numerical components for molecular and viscous flow, respectively.
This model relies exclusively on design and typical material parameters, together with measured values of intrinsic resonance frequency $f_\mathrm{in}$ and quality factor $Q_\mathrm{in}$. 
Measuring $f_\mathrm{in}$ and $Q_\mathrm{in}$ at a pressure $<\!10^{-7}\,\mathrm{mbar}$ self-calibrates our sensor over its entire measurement range.
For a trampoline's fundamental out-of-plane vibrational mode, the resulting deviation between measured and simulated pressure dependencies of the quality factor and resonance frequency is within $15\,\%$ and $4\,\%$, respectively. 
The resulting error for pressure values inferred from quality factor and frequency measurements is  $<10\,\%$, for pressures between $\sim 10^{-6}$ and $\sim 10^{-1}\,\mathrm{mbar}$, and $<25\,\%$ for the complete 10-decade measurement range. 
Exceptions are two outliers with increased measurement errors, which might be related to the limited accuracy of our commercial pressure gauge.
Based on investigations with helium, we demonstrate the potential for extending this sensing capability to other gases, thereby highlighting the practical use of our sensor.
\end{abstract}


\maketitle

\section*{Introduction}
Establishing and accurately measuring low gas pressures is a prerequisite for cutting-edge technologies, such as semiconductor fabrication, quantum computing, and fundamental science.
Typical operating pressures $P$ span ultra-high vacuum ($<\!10^{-8}\,\mathrm{mbar}$) to ambient pressure ($\sim\!10^3\,\mathrm{mbar}$), which currently requires two to three different sensors for a complete coverage~\cite{nakhosteen2016handbook,shirhatti2020broad}.
Depending on their measurement range, these devices rely on different physical phenomena, such as the transport of thermal energy by, or the ionization of, residual gas particles, or the pressure-induced deflection of a diaphragm~\cite{nakhosteen2016handbook}.
While the latter approach represents a direct pressure measurement, the former two enable only an indirect determination, which involves gas-species-dependent properties, such as the atomic mass or the specific heat.
Consequently,
the implementation of transition regions between individual sensors and a separate calibration for each type of gas is required.
This adds complexity and cost and can limit the achievable measurement accuracy.

Recent advances in nanofabrication have enabled the realization of compact absolute~\cite{shirhatti2020broad} and differential~\cite{kim2022self,chen2022nano} pressure sensors with wide measurement ranges, hinting at the possibility of covering ultra-high vacuum to ambient pressure with a single sensor.
However, realizing such a device still requires increasing measurement range and accuracy, compared to demonstrated sensors, and establishing a physical model or a calibration function, to describe the sensor's pressure dependency.
To this end, nano- and micro-mechanical resonators are promising candidates~\cite{eaton1997micromachined}.
A whole variety of these devices have been demonstrated with a sensitivity partially or fully covering the range $10^{-3}$ to $10^{3}\,\mathrm{mbar}$~\cite{kokubun1984bending, blom1992dependence, bianco2006silicon, li2007ultra, martin2007gas, southworth2009pressure, lubbe2011measurement, smith2013electromechanical, lee2014air, dolleman2016graphene, smith2016piezoresistive, naesby2017effects, wagner2018highly, alcheikh2019highly, song2020recent, ghatge202030}.
Furthermore, it has been pointed out that increasing the intrinsic quality factors $Q_\mathrm{in}$, e.g., by optimizing the device geometry, can significantly extend their measurement range towards lower pressure ranges~\cite{reinhardt2018ultralow}.
Recently, the sensitivity of a phononic crystal membrane resonator has been demonstrated in the range from $10^{-7}$ to $2\times10^{-5}\,\mathrm{mbar}$, enabled by its ultra-high $Q_\mathrm{in}\sim 10^8$~\cite{saarinen2022laser}.

The measurement principle of resonant mechanical sensors relies on a pressure-dependency of either their frequency $f_\mathrm{m}$, quality factor $Q$, or amplitude~\cite{christian1966theory, newell1968miniaturization, langdon1985resonator}.
In the free-molecular flow (FMF) regime, where the gas particles do not interact with each other, $Q$ follows~\cite{christian1966theory}
\begin{equation}
    Q=\bigl(Q_\mathrm{in}^{-1}+Q_\mathrm{FMF}^{-1}\bigr)^{-1}.  \label{eqn:Q_mod_1}
\end{equation}

Here, $Q_\mathrm{in}$ is the intrinsic quality factor of the resonator and $Q_\mathrm{FMF}$ describes the pressure dependency resulting from collisions with gas particles.
The latter is given by
\begin{equation}
    Q_\mathrm{FMF}=\frac{f_\mathrm{in}\rho h}{P}\sqrt{\frac{\pi^3k_\mathrm{B}T_\mathrm{gas}}{8m_\mathrm{gas}}}, \label{eqn:Q_FMF}
\end{equation}

with intrinsic resonance frequency $f_\mathrm{in}$, mass density $\rho$, and thickness $h$ of the resonator, as well as temperature $T_\mathrm{gas}$ and atomic mass $m_\mathrm{gas}$ of the gas.
When continuously increasing the pressure, the interaction between gas particles becomes important, such that the gas acts as a fluid.
This has two main consequences:
First, viscous damping is the dominating loss mechanism acting on the resonator.
Second, a fraction of the gas co-oscillates with the resonator, which contributes an additional load $m_\mathrm{add}$ to its intrinsic mass $m_\mathrm{eff}$, thereby reducing its oscillation frequency according to~\cite{sader1998frequency,yadykin2003added}

\begin{equation}
    f_\mathrm{m}=f_\mathrm{in}\left[1+m_\mathrm{add}/m_\mathrm{eff}\right]^{-1/2}. \label{eqn:f_visc}
\end{equation}

The specific pressure at which the transition between FMF and viscous flow (VF) regimes occurs depends on the properties of both device and fluid~\cite{verbridge2008size, kara2015nanofluidics}.
Simplified models, such as the two-dimensional (2D) oscillating cylinder model~\cite{sader1998frequency}, are used to describe viscous damping of cantilevers and beams\footnote{Here and in the following beams are considered to be doubly-clamped.}~\cite{ bhiladvala2004effect,ghatkesar2008resonating}.
Resonators with geometries corresponding to arrangements of multiple beams have been modelled by representing the device as a string of spheres~\cite{kokubun1984bending,hosaka1995damping}.
This approach makes use of the well-known analytical solution for a sphere oscillating in a viscous fluid~\cite{landau2013fluid}.
An approximate analytical model describing gas-induced loss over the entire range, from FMF to VF, and the intermediate transitional flow regime, has been established and effectively applied to beams and cantilevers~\cite{karabacak2007high, yakhot2007stokes}.
To overcome the limitations of analytical models, a recent article~\cite{liem2021nanoflows} describes a finite element method (FEM) for modelling $f_\mathrm{m}$ and $Q$ of cantilevers and beams in the VF regime.

Here, we investigate gas pressure sensing with ultra-high-$Q$ mechanical trampoline resonators~\cite{reinhardt2016ultralow, norte2016mechanical}, featuring $Q_\mathrm{in}\sim10^7$.
For the fundamental out-of-plane mode of the best device, we observe a continuous change in $Q$ from $Q=5\times10^6$ at $10^{-7}\,\mathrm{mbar}$ to $Q=7$ at $10^3\,\mathrm{mbar}$.
This corresponds to a sensitivity range of ten decades, which, to the best of our knowledge, is unprecedented.
Depending on the trampoline’s geometry, vibrational mode, and investigating either air or helium, a transition between FMF and VF occurs between $1$ and $100\,\mathrm{mbar}$.
We find that the model 
\begin{equation}
    Q_\mathrm{mod}=\bigl[Q_\mathrm{in}^{-1}+(Q_\mathrm{FMF}+Q_\mathrm{VF})^{-1}\bigr]^{-1}  \label{eqn:Q_mod_2}
\end{equation}
well-describes the observed behavior over the investigated range, where $Q_\mathrm{VF}$ represents the $P$-dependent $Q$ in the VF regime.
In a first step, we express $Q_\mathrm{VF}$ as a two-parameter fit function, which provides limited agreement with the data.
In a second step, to establish a more accurate model of viscous damping, we simulate the interaction of our device with the surrounding fluid via FEM in Comsol Multiphysics, similar to Ref. \cite{liem2021nanoflows}.
By substituting $Q_\mathrm{VF}$ in Eq.~\ref{eqn:Q_mod_2} with the simulated values, we find an overall agreement between data and combined analytical-FEM model within $\leq\pm15\,\%$.
The only model inputs are design and typical material parameters, together with $f_\mathrm{in}$ and $Q_\mathrm{in}$.
Self-calibration of our sensor, from near ultra-high vacuum to ambient pressure, is implemented by measuring $f_\mathrm{in}$ and $Q_\mathrm{in}$ at a pressure $<10^{-7}\,\mathrm{mbar}$.
This alleviates the need for a resource demanding calibration by comparison with a reference gauge \cite{ISO3567}.  
In addition to $Q$, we investigate the $P$-dependency of the trampolines' resonance frequencies $f_\mathrm{m}$ in the VF regime.
We find a significant reduction with respect to $f_\mathrm{in}$ of up to 38~\% at ambient pressure.
The resonance frequency simulated with our FEM model matches the data within $4\,\%$ over three decades, between $1$ and $1000\,\mathrm{mbar}$.
The error associated with pressure values inferred from $Q$ and $f_\mathrm{m}$ measurements is determined to be $<10\,\%$, for pressures between $\sim 10^{-6}$ and $\sim 10^{-1}\,\mathrm{mbar}$, and $<25\,\%$ between $\sim 10^{-7}$ and $\sim 10^{3}\,\mathrm{mbar}$. 
For two pressure values, lying between $1$ and $10\,\mathrm{mbar}$, we find an increased error of up to $57\,\%$.
We argue that the deviations between data and models for both $Q$ and $f_\mathrm{m}$, particularly the two outliers, might be dominated by a commercial pressure gauge.
For one of our devices, we also use helium gas, in addition to air, finding equally good agreement between data and fit functions.
Finally, we discuss possible routes for extending the measurement range towards ultra-high vacuum, for which currently only ionization gauges~\cite{jousten2020review} and cold atoms \cite{scherschligt2017development} provide adequate sensitivities.


\section*{Trampoline Pressure Sensor}

\subsection*{Measurement Setup}
\begin{figure}
\centering
\includegraphics[width=240pt]{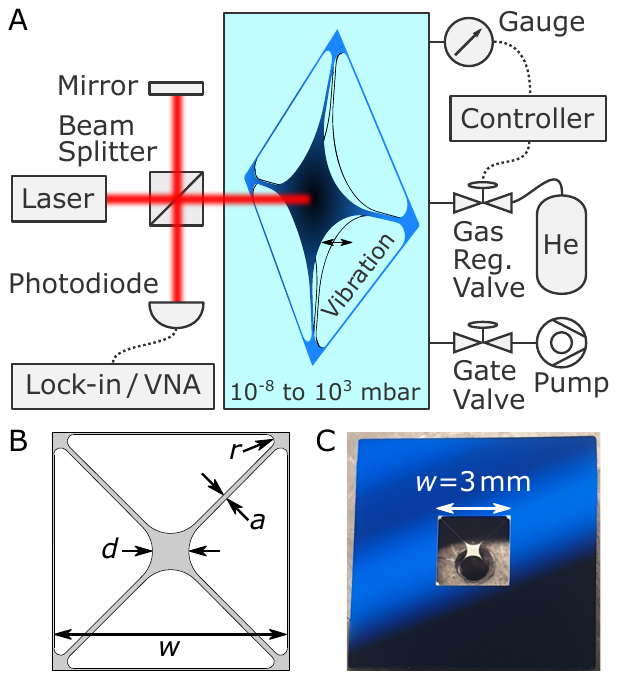}
\caption{
Pressure sensing with nanomechanical trampoline resonators.
A) A trampoline is situated inside an ultra-high-vacuum chamber.
Thermal energy in the environment or resonant excitation via a piezo actuator (not shown) cause the trampoline to vibrate at its characteristic resonance frequencies $f_\mathrm{m}$ (measured fundamental mode shown).
Surrounding gas impacts the vibration via a frictional force and, in the viscous flow regime, an increased effective mass.
The trampoline's vibration is measured with a Michelson interferometer
comprising a laser, a beam splitter, a mirror, and a photodiode connected to a lock-in amplifier or a vector network analyzer.
For our proof of principle, the pressure in the chamber is monitored with commercial gauges and adjusted through a gas regulating valve, both connected to a controller.
In combination with adjusting the effective pumping speed of a turbo pump, via a gate valve, the pressure in the chamber is set between $10^{-8}$ and $10^3$\,mbar.
(B)-(C) Trampoline design and investigated trampoline resonator with window size $w=3\,\mathrm{mm}$, pad diameter $d=500\,$\textmu m, tether width $a=2\,$\textmu m, and corner fillet radius $r=75\,$\textmu m.
}
\label{fig1}
\end{figure}

Figure~\ref{fig1}\textit{A} shows a schematic of the experimental setup used for investigating the trampoline pressure sensor.
Here, the device is placed inside an ultra-high vacuum chamber, in which the pressure is set between $10^{-8}$ and $10^3$\,mbar.
This is realized by injecting controlled amounts of gas (e.g., helium), via a gas-regulating valve (Pfeiffer EVR 116) and a corresponding controller (Pfeiffer RVC 300).
Furthermore, the opening of a gate valve, located in front of a turbo pump, is adjusted, to set the effective pumping speed.
The pressure inside the chamber is measured with a
Bayard-Alpert/Pirani combination gauge (Pfeiffer PBR 260), between $10^{-8}$ and $10$\,mbar, and a capacitive gauge (Pfeiffer CMR 361), between $10$ and $10^3$\,mbar.
To measure the oscillation amplitude for some of the trampoline's out-of-plane vibrational modes (fundamental mode is shown), a Michelson interferometer is used.
Here, an incident laser, having a wavelength of 1064\,nm, is split into two beams of equal intensity.
These beams are reflected of either a mirror or the trampoline and subsequently interfere at a photodiode (PD).
The PD's output signal, which is modulated at $f_\mathrm{m}$, is measured either with a lock-in amplifier or a spectrum analyzer.

Figure~\ref{fig1}\textit{B} shows a drawing of the trampoline design~\cite{reinhardt2016ultralow, norte2016mechanical}, where the device with lateral extent $w$ comprises a central pad of width $d$, which is suspended by four tethers of width $a$.
Corner fillets, defined via circular segments of radius $r$, connect the trampoline to the supporting silicon chip.
For the work presented in the following, we investigated two trampolines, which were obtained from Norcada Inc~\cite{norcada}.
They are made out of a silicon nitride thin film, having high tensile stress $\sigma=1\,\mathrm{GPa}$ and $\rho=3000\,\mathrm{kg/m^3}$, which is freely-suspended from a silicon chip.
The first device, referred to as trampoline i, is characterized by $w=1\,\mathrm{mm}$, $d=150\,$\textmu m, $a=21.5\,$\textmu m, $r=25\,$\textmu m, and $h=100\,\mathrm{nm}$.
The dimensions of the second device, referred to as trampoline ii, are $w=3\,\mathrm{mm}$, $d=500\,$\textmu m, $a=2\,$\textmu m, $r=75\,$\textmu m, and $h=80\,\mathrm{nm}$.
trampoline ii is shown in Fig.~\ref{fig1}\textit{C}, where the silicon nitride film appears blue on top of silicon and white in regions where it is freely-suspended.
Note that the given mass density and the specified thicknesses are used for all models presented in Fig.~\ref{fig3} to Fig.~\ref{fig5} (i.e., all oscillation modes of the same device use exactly the same parameters).
Possible variations in the membrane's thickness of $5\,\%$ and in its mass density of $10\,\%$ (see, e.g., Ref.~\cite{wilson2009cavity,saarinen2022laser}), might contribute to the systematic errors presented in Fig.~\ref{fig5}.

\subsection*{Sensor Readout}
\begin{figure}
\centering
\includegraphics[width=240pt]{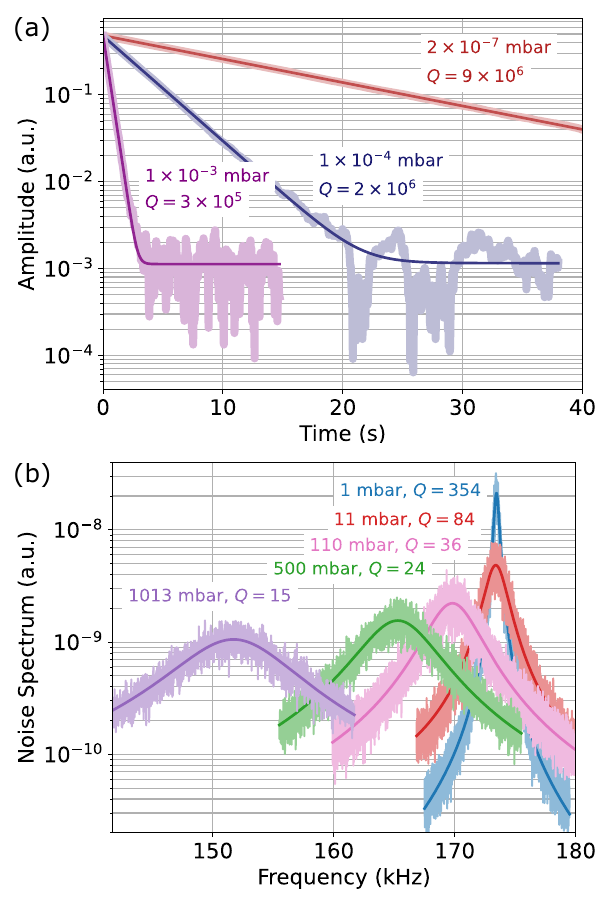}
\caption{
Measuring mechanical quality factor $Q$ and resonance frequency $f_\mathrm{m}$ at different air pressures $P$ (shown plots represent the fundamental mode of trampoline i, described in the main text.
For $P<0.1$\,mbar, the gas mainly impacts the resonator's $Q$.
Increasing $P$ above this level causes a reduction in $f_\mathrm{m}$, due to gas particles vibrating together with the resonator, thereby increasing its effective mass.
\textit{(A)} Typical ringdown of oscillation amplitude versus time $t$, occurring after switching of piezo-excitation, at three different pressure values.
$Q$ is extracted from the fit function $[(a_1e^{-\pi f_m t/Q})^2+b_1^2]^{-1/2}$ (solid line), with initial amplitude $a_1$ and thermal noise level $b_1$ ($Q$, $a_1$, and $b_1$ are fit parameters).
\textit{(B)} Typical thermal displacement noise spectra for five different pressure values.
For $P\gtrsim10^{-2}$\,mbar, $Q$ and $f_\mathrm{m}$ are extracted from fitting a Lorentzian function (solid line) $a_2\Bigl\{\left[4\pi^2(f_m^2-f^2)\right]^2+(4 \pi^2 f_m f/Q)^2\Bigl\}^{-2}+b_2$, with fit parameters $Q$, $f_\mathrm{m}$, $a_2$, and $b_2$.
}
\label{fig2}
\end{figure}

To measure a trampoline's $f_\mathrm{m}$ and $Q$, 
we perform mechanical ringdowns of its oscillation amplitude, for $P\lesssim 10^{-2}\,\mathrm{mbar}$.
At higher pressures, due to the ringdown's short duration ($\lesssim 0.1\,\mathrm{s}$ for the oscillation amplitude to decay by a factor $1/e$), we assess the noise spectra related to thermally-driven oscillations.
A mechanical ringdown is obtained by resonantly exciting an oscillation mode via a piezo actuator, which is connected to the lock-in amplifier's output (we have used piezos placed either in- or outside the UHV chamber; effective excitation with the external piezo, which is mounted to the same optical table as the trampoline, is facilitated by the trampolines' exquisite force sensitivity \cite{reinhardt2016ultralow}).
Upon switching off the piezo drive, the trampoline's oscillation amplitude follows an exponential decay, until it reaches the thermal noise floor, where it undergoes random (i.e., Brownian) motion, as shown in Fig.~\ref{fig2}\textit{A}.
Here, data sets for trampoline i's fundamental oscillation mode at air pressures of $2\times10^{-7}\,\mathrm{mbar}$, $10^{-4}\, \mathrm{mbar}$, and $10^{-3}\,\mathrm{mbar}$ are displayed, corresponding to quality factors of $9\times10^{6}$, $2\times10^{6}$, and $3\times10^{5}$, respectively.
The $Q$ values are obtained from fitting $[(a_1e^{-\pi f_\mathrm{m} t/Q})^2+b_1^2]^{-1/2}$ (lines in Fig.~\ref{fig2}\textit{A}), with fit parameters $a_1$, $a_2$, time $t$, and intrinsic resonance frequency $f_\mathrm{in}=173.6\,\mathrm{kHz}$, to the data~\cite{reinhardt2016ultralow}. 
Here, $\tau=Q/ (\pi f_\mathrm{in})$ corresponds to the ringdown time. 

Figure~\ref{fig2}\textit{B} shows displacement noise spectra of trampoline i's fundamental out of plane mode for five different air pressures.
$Q$ and $f_{m}$ are obtained from fitting a Lorentzian (lines in Fig.~\ref{fig2}\textit{B}) $a_2\Bigl\{\left[4\pi^2(f_m^2-f^2)\right]^2+(4 \pi^2 f_m f/Q)^2\Bigl\}^{-2}+b_2$, with fit parameters $a_2$, and $b_2$, to the data.
Up to $\sim10\,\mathrm{mbar}$, the effect of increasing $P$ is limited to broadening the resonance, thereby decreasing $Q$.
Further raising $P$ additionally causes a reduction in the resonance frequency, thereby indicating the FMF-VF transition, consistent with Eq.~\ref{eqn:f_visc}.
At ambient pressure ($1013\,\mathrm{mbar}$), $f_\mathrm{in}-f_\mathrm{m}=21\,\mathrm{kHz}$, which corresponds to 13~\%.

The time it takes to obtain a pressure reading with our sensor is $\sim\tau$, for ringdown measurements, and $\sim100\tau$ for spectral measurements. 
In the longest case, corresponding to trampoline ii at base pressure $\sim10^{-7}\,\mathrm{mbar}$, the measurement time is $\sim100\,\mathrm{s}$.

\subsection*{Pressure Dependency of $Q$ and $f_\mathrm{m}$}
\begin{figure*}\includegraphics[width=504pt]{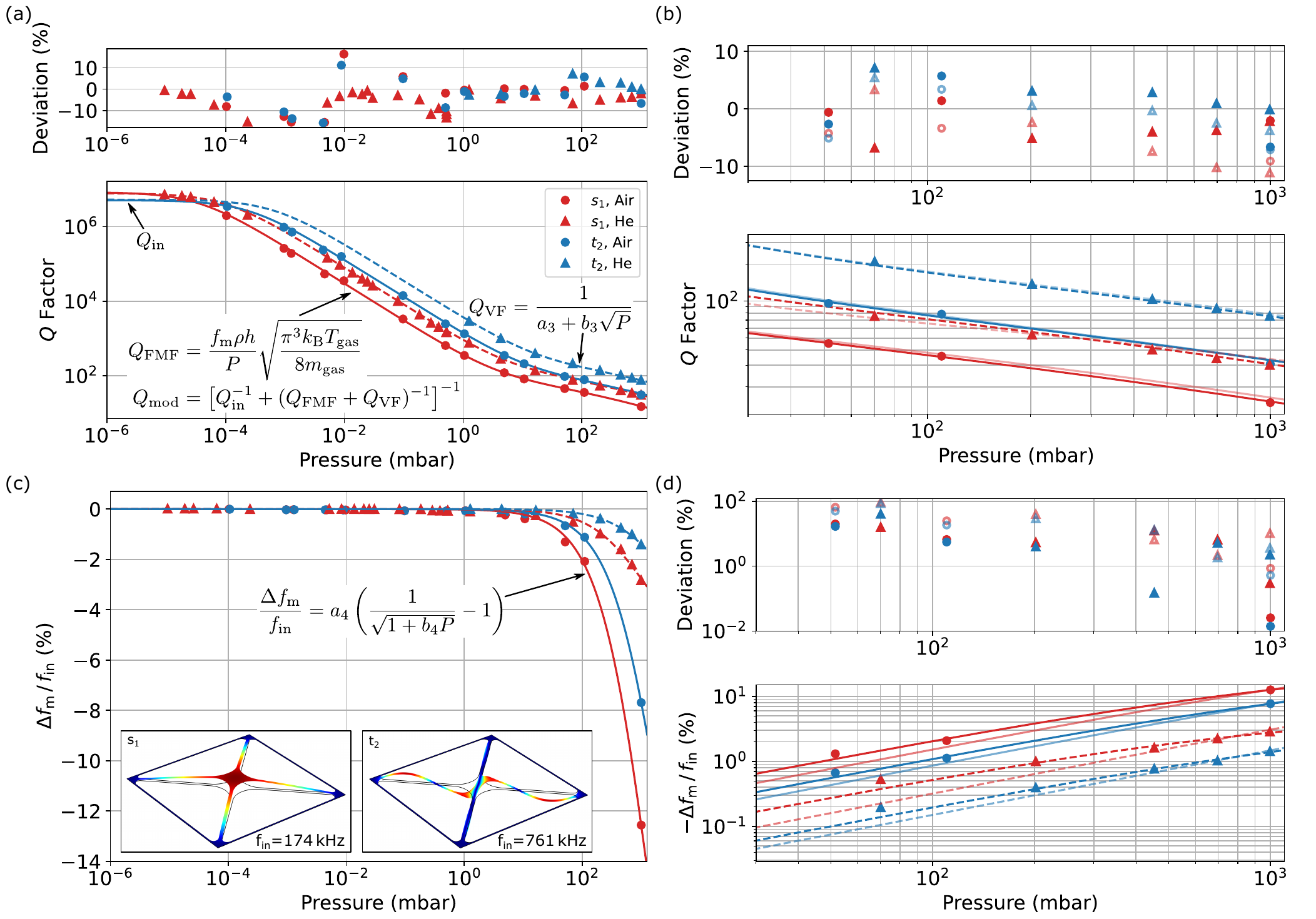}
\caption{Dependency of mechanical quality factor $Q$ and resonance frequency $f_\mathrm{m}$ on air and helium pressure for trampoline i.
\textit{(A, lower)} $Q$ as a function of pressure for s\textsubscript{1} (red) and t\textsubscript{2} (blue) modes (see insets of panel (c) for corresponding mode profiles (simulated with Comsol) and intrinsic resonance frequencies $f_\mathrm{in}$ (measured)).
Measured values $Q_i$ (average of four to seven individual measurements) are represented by circles and triangles for air and helium, respectively.
Solid and dashed curves show corresponding model functions $Q_\mathrm{mod}(P)$, with intrinsic quality factor $Q_\mathrm{in}$, free-molecular flow model $Q_\mathrm{FMF}$, and viscous flow fit function $Q_\mathrm{VF}$.
Explicit expressions are given in the plot, with mass density $\rho=3000\,\mathrm{kg/m^3}$ and thickness $h=100\,\mathrm{nm}$ of trampoline i, temperature $T_\mathrm{gas}=293\,\mathrm{K}$ and masses $m_\mathrm{gas}=\{4.81,0.66\}\times10^{-27}\,\mathrm{kg}$ of $\{\mathrm{air},\mathrm{He}\}$ particles, and fit parameters $a_3$ and $b_3$.
The transition between free-molecular and viscous flow regimes occurs between $\sim1\,\mathrm{mbar}$~and~$\sim100\,\mathrm{mbar}$, depending on mode and gas species.
(A, upper) Relative deviation of measured values and model function $Q_i/Q_\mathrm{mod}(P_i)-1$.
\textit{(B)} Detail of \textit{(A)}, showing only the viscous flow regime.
The light red and blue curves (lower) represent an analytical model for viscous damping of an oscillating string of spheres, where the sphere radius is a fit parameter (see main text for details).
The corresponding deviation with respect to the measured values is represented by light open markers (upper).
\textit{(C)} Relative change in frequency, $\Delta f_\mathrm{m}/f_\mathrm{in}=f_\mathrm{m}/f_\mathrm{in}-1$, as a function of $P$.
Solid and dashed curves represent fit functions (expression given in the plot), with fit parameters $a_4$ and $b_4$.
\textit{(D, lower)} Detail of \textit{(C)}, showing only the viscous flow regime.
The light red and blue curves (lower) represent an analytical model for viscous damping of a cantilever, with a single size- and mode-shape-dependent fit parameter (see main text for details).
\textit{(D, upper)} Solid (open) markers represent deviations between measured values and the two-parameter fit function,
    given in \textit{(C)}, (between measured values and the cantilever model).
}\label{fig3}\end{figure*}

Figure~\ref{fig3}\textit{A} (lower) shows $Q$ as a function of air or helium pressure for the first symmetric (s$_1$) and the second torsional (t$_2$) mode of trampoline i.
Related mode profiles and intrinsic resonance frequencies, simulated in Comsol, are shown in the insets of Fig.~\ref{fig3}\textit{C}.
Measured values for s$_1$ (t$_2$) are represented by red (blue) dots and triangles for air and helium, respectively.
Each $Q$ value is obtained by averaging four to seven individual measurements.
The corresponding uncertainties are typically few percent, where fitting errors appear to be a main limitation.
This has been verified, by fitting different sections of ringdown curves and noise spectra.
In the best scenario, a standard deviation of $0.5\,\%$ has been obtained, for a set of six subsequent ringdowns taken at base pressure $P_\mathrm{base}=5.4\times10^{-8}\,\mathrm{mbar}$.
Here, for each ringdown, the trampoline was driven to exactly the same oscillation amplitude, exactly when the previous decay reached the onset of the thermal noise floor.
This indicates the potential for high-precision pressure measurements with nanomechanical oscillators.
For other ringdown measurements, showing larger deviations, the maximum oscillation amplitude was not well controlled and part of the thermal noise floor was included in the fit.
The model function $Q_\mathrm{mod}$, corresponding to each data set, is given by Eq.~\ref{eqn:Q_mod_2}.
They are shown as solid and dashed curves for air and He, respectively.
Here, $Q_\mathrm{in}$ corresponds to the quality factor measured at $P_\mathrm{base}$, which is $8.0\times10^6$ and $5.1\times10^6$ for s$_1$ and t$_2$, respectively.
These values are about $30\,\%$ lower than the ones simulated in Comsol, for a limitation due to bending loss \cite{yu2012control,tsaturyan2017ultracoherent}, under the assumption of a typical material quality factor of $5000$ for a $100\,\mathrm{nm}$ thick SiN membrane \cite{villanueva2014evidence}. 
This discrepancy is consistent with studies of similar devices \cite{hoj2021ultra}, which point out  additional loss due to radiation of elastic waves to the silicon chip as possible cause.
$Q_\mathrm{FMF}$ is given by Eq.~\ref{eqn:Q_FMF}, with  $T_\mathrm{gas}=293\,\mathrm{K}$ and
$m_\mathrm{gas}=\{4.81,0.66\}\times10^{-27}\,\mathrm{kg}$ for $\{\mathrm{air},\mathrm{helium}\}$ particles.
The VF component is represented by a fit function, $Q_\mathrm{VF}=(a_3+b_3\sqrt{P})^{-1}$, with fit parameters $a_3$ and $b_3$.
It represents a generalized version of the expression for a viscously damped cantilever, assumed to consist of a string of spheres (SOS), which is given by~\cite{lubbe2011measurement, hosaka1995damping} $Q_\mathrm{SOS}=4\rho h D f_\mathrm{in}/\left[6\mu+3D\sqrt{\pi\mu(m_\mathrm{gas}/k_\mathrm{B}T)f_\mathrm{in}P}\right]$, with dynamic viscosity of the gas $\mu$ and sphere diameter $D$.

The onset of the pressure dependency occurs from $10^{-5}\,\mathrm{mbar}$ to $10^{-3}\,\mathrm{mbar}$ and the transition between molecular and viscous flow regimes in the range from $1\,\mathrm{mbar}$ to $100\,\mathrm{mbar}$, both depending on resonance mode and gas species.
Our model corroborates this dependency as both transitions are determined by the membrane properties $\rho$, $h$, and $f_\mathrm{in}$, together with gas parameters $m_\mathrm{gas}$ and $T$.
Furthermore, increasing $Q_\mathrm{in}$ and $D$ moves the onset of $Q$'s pressure dependency and the FMF-VF transition to lower pressures, respectively.
While molecular damping (Eq.~\ref{eqn:Q_FMF}) depends only on the thickness of the resonator, viscous damping ($Q_\mathrm{SOS}$) also involves the lateral extent, represented by $D$.
This hints at a dependency on the geometry of viscous damping affecting a thin-film resonator, as observed in~\cite{verbridge2008size}.

Figure~\ref{fig3}\textit{A} (upper) shows the relative deviation $Q_i/Q_\mathrm{mod}(P_i)-1$ between data points $Q_i$, measured at pressures $P_i$, and corresponding model values.
Overall, data and model agree within $\pm15\,\%$, which matches the specified accuracy of our pressure gauge~\cite{PfeifferPBR260}; between $1\,\mathrm{mbar}$ and $10^3\,\mathrm{mbar}$, the deviation is within $\pm10\,\%$.
The similarity in the deviations of both modes obtained with air in the molecular flow regime hints at the inaccuracy of the commercial pressure gauge as probable cause for the deviations.
In the VF regime, the semiempirical fit function might pose a significant limitation to the attainable accuracy.
Figure~\ref{fig3}\textit{B} (lower) shows the corresponding pressure range, where in addition to $Q_\mathrm{mod}$ a second model function is shown, in light red and light blue for air and He, respectively.
In this function (Eq.~\ref{eqn:Q_mod_2}), $Q_\mathrm{VF}$, is replaced by $Q_\mathrm{SOS}$, leaving $D$ as the sole fit parameter.
Corresponding values of $D_\mathrm{s_1}=45\,$\textmu m and $D_\mathrm{t_2}=23\,$\textmu m are obtained for s$_1$ and t$_2$, respectively, which lie within trampoline i's tether and pad widths (see above).
$D_\mathrm{s_1}>D_\mathrm{t_2}$ is consistent with the larger contribution of the pad to the modes' displacement profiles, for s$_1$ compared to t$_2$.
The deviations between data and $Q_\mathrm{SOS}$ (light red/blue open symbols) exceed $10\,\%$, in the vicinity of ambient pressure for t$_2$, as shown in Fig.~\ref{fig3}\textit{B} (upper), thereby showcasing the advantage of $Q_\mathrm{VF}$, which yields deviations $<10\,\%$.

Figure~\ref{fig3}\textit{C} shows the relative change in oscillation frequency $\Delta f_\mathrm{m}/f_\mathrm{in}$, with $\Delta f_\mathrm{m}=f_\mathrm{m}-f_\mathrm{in}$, over the investigated pressure range.
Data sets are shown together with corresponding fit functions $a_4\left[\left(1+b_4P\right)^{-1/2}-1\right]$, having fit parameters $a_4$ and $b_4$.
This function is generalized from the expression for a cantilever~\cite{sader1998frequency}, for which $a_4=1$ and $b_4$ depends on oscillation mode and gas properties.
The plotting style for both modes and gas types is identical to Fig.~\ref{fig3}\textit{A}.
With regard to Fig.~\ref{fig3}\textit{A}, it is apparent that a significant decrease in $f_\mathrm{m}$ only occurs in the VF regime.
The onset of the effect agrees with the FMF-VF transition.
Furthermore, it is more pronounced for s$_1$ compared to t$_2$, which implies a correspondingly bigger fluid load $m_\mathrm{add}$, according to Eq.~\ref{eqn:f_visc};
in particular, since $m_\mathrm{eff}$ of s$_1$ is larger compared to t$_2$~\cite{reinhardt2016ultralow}.
The frequency decrease is significantly more pronounced for air compared to He, which is consistent with the corresponding difference in their particle masses.

A detail of the frequency decrease (Fig.~\ref{fig3}\textit{C}), covering the VF regime, is shown in Fig.~\ref{fig3}\textit{D}.
In addition to the two-parameter fits, this plot also contains fits with $a_4=1$ (i.e., of a cantilever type), shown in light red and light blue for air and He, respectively.
Corresponding deviations are shown in Fig.~\ref{fig3}\textit{D} (upper), where light colored open symbols represent the deviation between data and cantilever-type fits.
These deviations exceed the ones obtained with two free fit parameters multiple times, for most data points.
Between $100\,\mathrm{mbar}$ and $1000\,\mathrm{mbar}$, the two-parameter fit matches the data within $10\,\%$, while at lower pressures the deviations are larger.

\subsection*{Self-calibrated Pressure Sensing}
\begin{figure*}
\centering
\includegraphics[width=469.7pt]{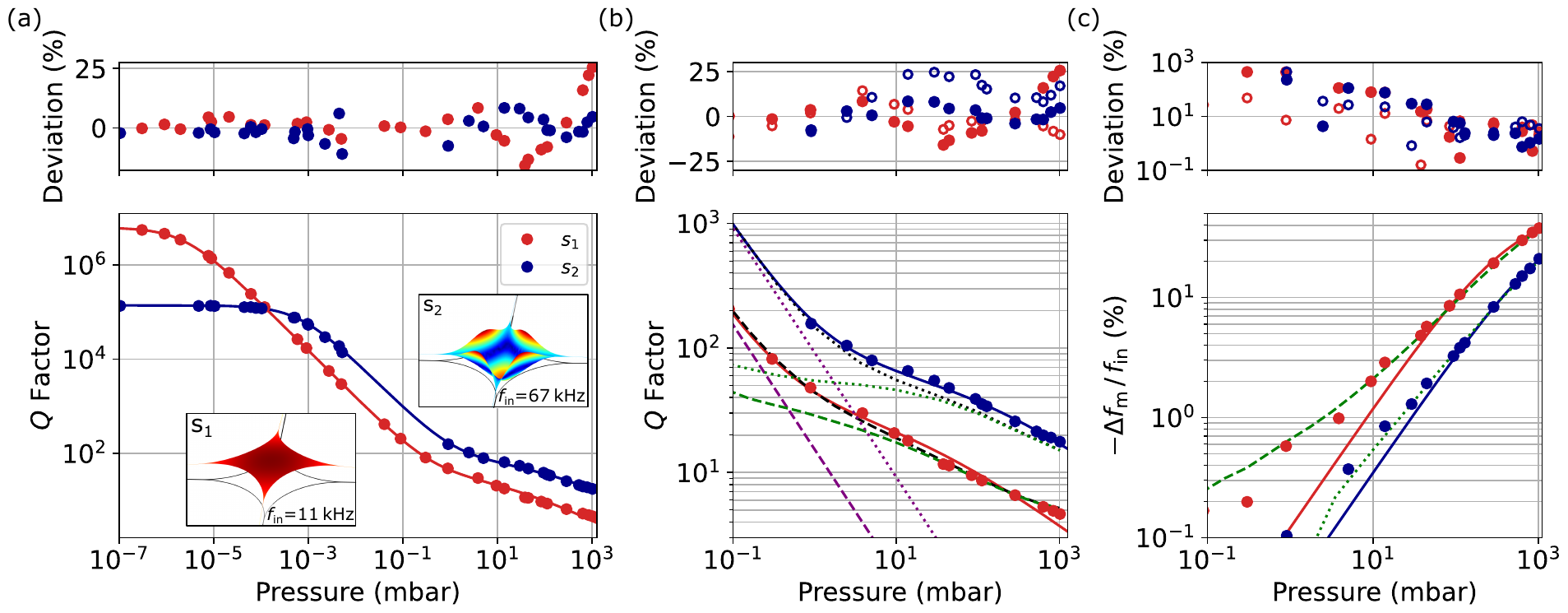}
\caption{
Self-calibrating pressure sensing over ten decades of air pressure with trampoline ii.
\textit{(A)} Similar to Fig.~\ref{fig3}\textit{A}: The lower panel shows $Q$ VS $P$ for s\textsubscript{1} (red) and s\textsubscript{2} (dark blue) modes, with simulated mode profiles (Comsol) and intrinsic resonance frequencies (measured) shown in the insets.
The solid lines represent semiempirical model functions (Eq.~\ref{eqn:Q_mod_2}; parameters given in the main text), which include two fit parameters.
Deviations between measured values and model functions are shown in the upper panel.
\textit{(B)} Detail of \textit{(A)} showing the upper four pressure decades.
Purple, green, and black curves show the analytical model for free-molecular flow , $Q_\mathrm{FMF}$, the Comsol model for viscous flow, $Q_\mathrm{COM}$, and the combined model, $\Tilde{Q}_\mathrm{mod}(P)=\bigl[Q_\mathrm{in}^{-1}+(Q_\mathrm{FMF}+Q_\mathrm{COM})^{-1}\bigr]^{-1}$, which rely exclusively on known membrane and gas parameters.
Here, dashed (dotted) curves are for s\textsubscript{1} (s\textsubscript{2}).
Open markers in the upper panel show deviations between measured values and $\Tilde{Q}_\mathrm{mod}$, which are within $15\,\%$ for s\textsubscript{1} over the investigated range.
\textit{(C)} Similar to Fig. \ref{fig3}\textit{D}: Relative change in $f_\mathrm{m}$ as a function of $P$.
Solid curves represent fit functions (expression given in Fig.~\ref{fig3}\textit{C}).
The green curves are obtained with Comsol.}
\label{fig4}
\end{figure*}

Figure~\ref{fig4}\textit{A} (lower) shows the dependency on air pressure of $Q$ for trampoline ii (see Fig.~\ref{fig1}\textit{C}).
The first and the second symmetric mode, designated as s$_1$ and s$_2$, with $Q_\mathrm{in}$ of $6.1\times10^6$ and $1.4\times10^5$, respectively, have been investigated;
corresponding mode profiles and intrinsic resonance frequencies, simulated in Comsol, are shown in the insets.
Measured values of $Q_\mathrm{in}$ are lower by a factor 3 compared to the ones simulated in Comsol, for a limitation due to bending loss \cite{yu2012control,tsaturyan2017ultracoherent}. 
Here, a typical material quality factor for an $80\,\mathrm{nm}$ thick SiN membrane of $4000$ \cite{villanueva2014evidence} is assumed.
The discrepancy between measured and simulated $Q$ factors is similar to the one reported in Ref.~\cite{bereyhi2022hierarchical}. 
Here, the authors point out a potential relation between the increased loss and a static out-of-plane deformation (i.e., \textit{buckling}) of the device. 
In fact, we see evidence of buckling for trampoline ii, when viewing it under a microscope. 
Optimizing the device geometry might enable avoiding this limitation. 
Together with the data (dots), associated model functions (lines), given by Eq.~\ref{eqn:Q_mod_2}, with two-parameter fit function $Q_\mathrm{VF}$, are presented.
s$_1$ shows a pronounced pressure dependency from $\sim10^{-7}\,\mathrm{mbar}$ to $\sim10^{3}\,\mathrm{mbar}$, thereby realizing a 10-decade measurement range, which, to the best of our knowledge, is unprecedented.
The extended sensitivity towards lower pressure, compared to trampoline i, is a consequence of trampoline ii's sixteenfold lower resonance frequency.
Effects on the sensitivity range due to differences in $h$ and $Q_\mathrm{in}$, between trampoline i and ii, cancel each other within 14~\%.
The FMF-VF transition occurs between $0.1\,\mathrm{mbar}$ and $1\,\mathrm{mbar}$, i.e., at about tenfold lower pressure compared to s$_1$ of trampoline i.
This appears to be a consequence of trampoline ii's lower resonance frequency and larger spatial extent, compared to trampoline i, which follows from equating $Q_\mathrm{FMF}$ with the two-parameter fit function $Q_\mathrm{VF}$ and solving for $P$.
For s$_2$, a pressure dependency only from $\sim10^{-4}\,\mathrm{mbar}$, due to its comparatively low $Q_\mathrm{in}$, up to $\sim10^{3}\,\mathrm{mbar}$ is evident.
The FMF-VF transition of s$_2$ occurs between $0.4\,\mathrm{mbar}$ and $40\,\mathrm{mbar}$.
Deviations between data and model functions are shown in Fig.~\ref{fig4}\textit{A} (upper).
They are~$<15\,\%$ for s$_2$ over the investigated pressure range.
s$_1$ shows a divergence between data and model for $P>15\,\mathrm{mbar}$, which reaches up to $25\,\%$ at ambient pressure, thereby indicating a systematic deviation from the semiempirical model function $Q_\mathrm{VF}$.

To establish a more accurate model of viscous damping acting on the trampoline,
we have implemented a corresponding simulation in Comsol, by following the approach presented in Ref.~\cite{liem2021nanoflows} for cantilevers and beams.
Here, the trampoline's oscillation, implemented in the structural mechanics module, is coupled to the viscous fluid (governed by Navier-Stokes equations), implemented in the acoustics module.
The resulting model includes energy dissipation from the trampoline via both viscous drag and thermal conduction, where the latter is negligible, as temperature gradients are insignificant in our case.
Figure~\ref{fig4}\textit{B} shows a detail of Fig.~\ref{fig4}\textit{A}, focusing on the upper four decades of pressure.
Here, the simulated quality factors related to viscous damping, $Q_\mathrm{COM}$, are shown in green.
For s$_1$ and $P\geq40\,\mathrm{mbar}$, the simulation gives an excellent agreement with measured values.
By replacing $Q_\mathrm{VF}$ in Eq.~\ref{eqn:Q_mod_2} with $Q_\mathrm{COM}$, we establish a model, $\Tilde{Q}_\mathrm{mod}=\bigl[Q_\mathrm{in}^{-1}+(Q_\mathrm{FMF}+Q_\mathrm{COM})^{-1}\bigr]^{-1}$, which covers the investigated pressure range from FMF to VF and relies exclusively on known membrane parameters and gas type, together with $f_\mathrm{in}$ and $Q_\mathrm{in}$.
Measuring $f_\mathrm{in}$ and $Q_\mathrm{in}$ at $P<10^{-7}\,\mathrm{mbar}$ self-calibrates the sensor over its entire measurement range.  
$\Tilde{Q}_\mathrm{mod}$ is represented by the black line, which agrees with the data within $15\,\%$ over the entire measurement range, indicated by open red circles in Fig.~\ref{fig4}\textit{B} (upper).
Consequently, $\Tilde{Q}_\mathrm{mod}$ accurately predicts the pressure range over which the FMF-VF transition occurs.
For decreasing pressure, $\Tilde{Q}_\mathrm{mod}$ asymptotically converges to ${Q}_\mathrm{FMF}$ (purple).
For s$_2$, $Q_\mathrm{COM}$ provides systematically larger values (up to $25\,\%$), compared to the data.
This is considered to follow from marginal geometric imperfections, such as slight out-of-plane buckling of the central pad's rim (indicated in investigations with an optical microscope; could be avoided by a slight design optimization), which are not accounted for in the simulation.

Figure~\ref{fig4}\textit{C} shows $-\Delta f_\mathrm{m}/f_\mathrm{in}$ for trampoline ii in the VF regime.
Measured values (dots) are shown together with two-parameter fit functions (red and blue lines; see previous sec. for details) and $f_\mathrm{COM}/f_\mathrm{COM,in}-1$ (green dashed/dotted lines). 
Here, $f_\mathrm{COM}$ and $f_\mathrm{COM,in}$ are the values simulated in Comsol for the $P$-dependent resonance frequency in the VF regime and the intrinsic resonance frequency, respectively.
For s$_1$ and s$_2$ the oscillation frequency is reduced by $38\,\%$ and $21\,\%$, respectively, at ambient pressure.
From $10^2\,\mathrm{mbar}$ to $10^3\,\mathrm{mbar}$, the fit functions match the data within $10\,\%$.
However, for $P<10\,\mathrm{mbar}$, they significantly diverge from measured values, thereby ceasing to provide an accurate description.
The Comsol model matches the data over a significantly broader range, providing a deviation within $15\,\%$ for s$_1$, from $1\,\mathrm{mbar}$ to $10^3\,\mathrm{mbar}$, and within $25\,\%$ for s$_2$, from $10\,\mathrm{mbar}$ to $10^3\,\mathrm{mbar}$.

\subsection*{Measurement errors}
In the following, we assess the error of the pressure measurement obtained with the s\textsubscript{1} mode of trampoline ii (see Fig.~\ref{fig4}). 

\begin{figure*}
\centering
\includegraphics[width=469.7pt]{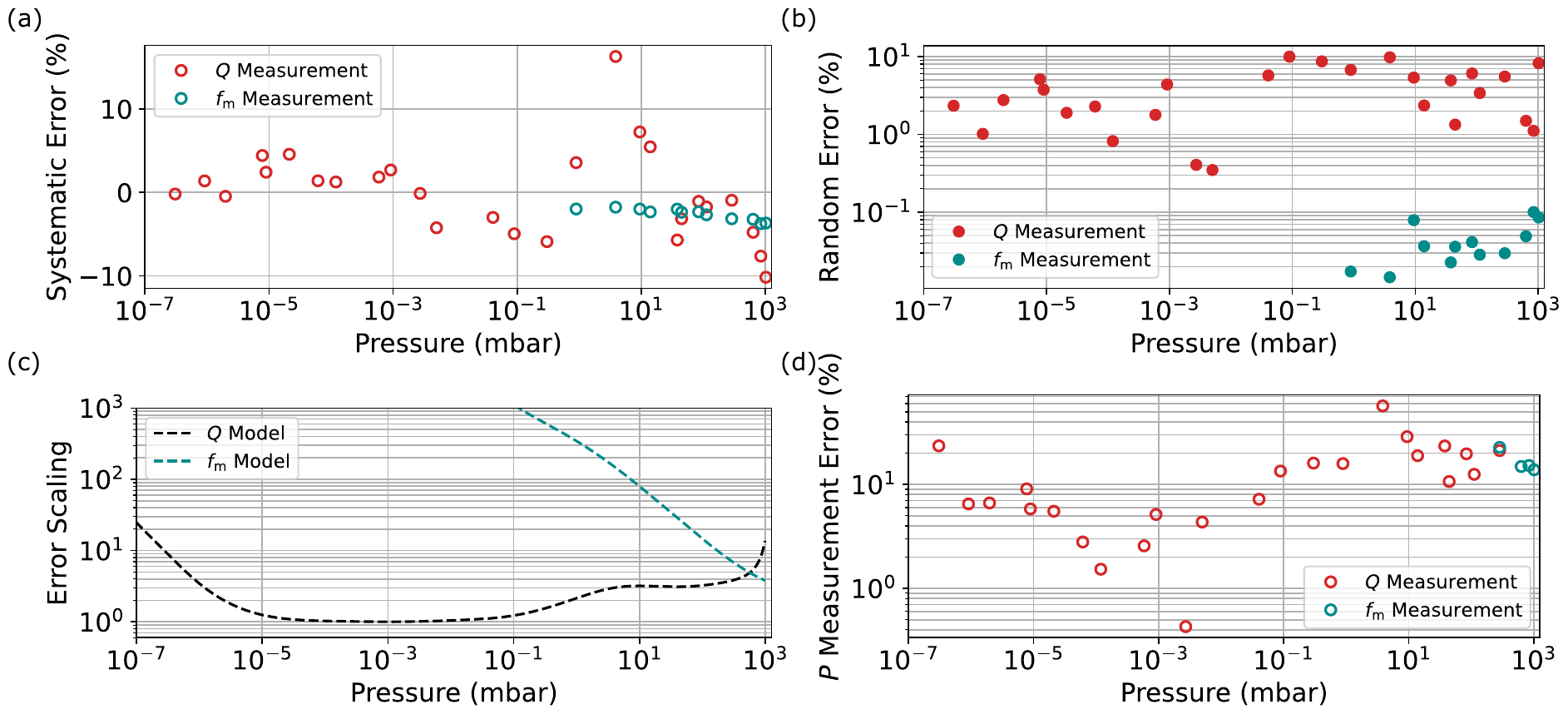}
\caption{
Errors of the air pressure measurement with the s\textsubscript{1} mode of trampoline ii.
(a) Systematic errors of measured quality factors $Q$ (red) and resonance frequencies $f_\mathrm{m}$ (cyan), corresponding to the relative deviations between data $\{Q_i,\,f_{\mathrm{m},i}\}$ and respective models $\{\Tilde{Q}_\mathrm{mod},\,f_\mathrm{COM}\}$ (see Fig.~\ref{fig4}). 
(b) Random errors of $Q$ and $f_\mathrm{m}$ measurements; each data point corresponds to the standard deviation of a set of repeated measurements.
(c) Proportionality constants relating the errors in $Q$ and $f_\mathrm{m}$ measurements, respectively, to inferred $P$ values (derived based on the model functions shown in Fig.~\ref{fig4} (b) and (c)). 
(d) Error of the $P$ values inferred from $Q$ and $f_\mathrm{m}$ measurements, corresponding to the root sum square of systematic and random errors (shown in panel (a) and (b), respectively) rescaled by the proportionality constant shown in panel (c).
Errors related to the $Q$ and $f_\mathrm{m}$ measurements are shown for $P\leq 287\,\mathrm{mbar}$ and $P> 287\,\mathrm{mbar}$, respectively, where they provide the lowest measurement error.
}
\label{fig5}
\end{figure*}

Figure~\ref{fig5}(a) shows the systematic errors, which are given by the relative deviations $\sigma_{\mathrm{S},Q}=Q_i/\Tilde{Q}_\mathrm{mod}(P_i)-1$ and $\sigma_{\mathrm{S},f}=f_{\mathrm{m},i}/f_\mathrm{COM}(P_i)-1$.
$\sigma_{\mathrm{S},Q}$ is within $\pm10\,\%$, except for one outlier at $3.9\,\mathrm{mbar}$, which might originate from an increased measurement error of $35\,\%$ of our commercial pressure gauge.
Under this assumption (i.e., the actual pressure being $2.5\,\mathrm{mbar}$ instead of $3.9\,\mathrm{mbar}$), the corresponding data points, both for $Q$ and $f_\mathrm{m}$ (red dots in Fig.~\ref{fig4}(b) and (c), respectively), would agree with respective model functions $\Tilde{Q}_\mathrm{mod}$ and $f_\mathrm{COM}/f_\mathrm{COM,in}-1$ (black dashed curve in Fig.~\ref{fig4}(b) and green dashed curve in Fig.~\ref{fig4}(c), respectively).    
The systematic frequency error is shown in the pressure range of the VF regime, where $f_\mathrm{m}$ shows a pronounced $P$-dependency.
Its absolute value reaches a minimum of about $2\,\%$, at the FMF-VF transition; incomplete reflection of the trampoline's geometric imperfections and material properties in the Comsol simulation are considered to be limiting factors.
Also, temperature or pressure variations in the sensor’s vicinity might contribute.
The independence of measured $Q_\mathrm{in}$ and $f_\mathrm{in}$ values from the incident optical power indicates no effects from laser-induced heating or dmaping. 

Random errors $\sigma_{\mathrm{S},Q}$ and $\sigma_{\mathrm{R},f}$ are expressed as the standard deviation of repeated $Q$ and $f_\mathrm{m}$ measurements, respectively, which are shown in Fig.~\ref{fig5}(b).
Values of $\sigma_{\mathrm{S},Q}$ obtained from ringdown measurements ($P<10^{-2}\,\mathrm{mbar}$) are on average about twofold smaller than the values extracted from fits of the thermal noise spectrum ($P>10^{-2}\,\mathrm{mbar}$).
The random error of the $f_\mathrm{m}$ measurement, shown in the VF pressure range, is on average about hundredfold smaller than the one of the $Q$ measurement.

While for the $Q$ measurement systematic and random errors contribute about equally, is the $f$ measurement dominated by systematic errors. 

To determine the overall error of the pressure values inferred from $Q$ and $f_\mathrm{m}$ measurements, the following expressions are applied, respectively:
\begin{align*}
\centering
\left(\frac{\Delta P}{P}\right)_Q &= \left(\frac{\partial \Tilde{Q}_\mathrm{mod}}{\partial P}\right)^{-1} \frac{\Tilde{Q}_\mathrm{mod}}{P}\sqrt{\sigma_{\mathrm{S},Q}^2+\sigma_{\mathrm{R},Q}^2}, \\    
\left(\frac{\Delta P}{P}\right)_f &=\left(\frac{\partial f_\mathrm{COM}}{\partial P}\right)^{-1} \frac{f_\mathrm{COM}}{P}\sqrt{\sigma_{\mathrm{S},f}^2+\sigma_{\mathrm{R},f}^2}.
\end{align*}
Here, the root sum square of systematic and random errors is scaled based on the corresponding model function and its derivative with respect to pressure. 
These scalings are shown in Fig.~\ref{fig5}(c), where, for the $Q$ measurement, five distinct ranges can be identified:
Between $10^{-7}\,\mathrm{mbar}$ and $10^{-5}\,\mathrm{mbar}$, where $\Tilde{Q}_\mathrm{mod}$ transitions from $Q_\mathrm{in}$ (constant) to $Q_\mathrm{FMF}\propto1/P$, the scaling reduces from 25 towards 1. 
Between $10^{-5}\,\mathrm{mbar}$ and $10^{-1}\,\mathrm{mbar}$, $\Tilde{Q}_\mathrm{mod}\approx Q_\mathrm{FMF}$, which results in a scaling of 1. 
In the FMF-VF transition region, occurring approximately between $10^{-1}\,\mathrm{mbar}$ and $5\,\mathrm{mbar}$, the scaling increases up to 3. 
It stays constant, up to about $10^2\,\mathrm{mbar}$, where it starts further increasing towards ambient pressure. 
This increase is due to the flattening out of $\Tilde{Q}_\mathrm{mod}$ (see Fig.~\ref{fig4}(b)), which might be linked to $Q$ approaching unity. 
The reasoning behind this assumption is, that neither for the s\textsubscript{2} mode of trampoline ii (Fig.~\ref{fig4}(b)) nor for the s\textsubscript{1} or t\textsubscript{2} modes of trampoline i (see Fig.~\ref{fig3}(a)) this behavior is evident. 
For these modes, $Q>10$ at ambient pressure. 
The error scaling related to $f_\mathrm{m}$ exceeds the one related to $Q$ for most of the investigated pressure range. 
This reflects a comparatively reduced pressure sensitivity of $f_\mathrm{m}$ compared to $Q$. 

Figure~\ref{fig5}(d) shows the errors associated with pressure values inferred from $Q$ and $f_\mathrm{m}$ measurements, according to the expressions given above.
The $Q$ measurement results in lower errors for $P\leq 287\,\mathrm{mbar}$, whereas $f_\mathrm{m}$ yields lower errors for $P>287\,\mathrm{mbar}$.  
Over a measurement range of five decades, $10^{-6}\,\mathrm{mbar}\lesssim P \lesssim 10^{-1}\,\mathrm{mbar}$, $\Delta P/P<10\,\%$.
At the low pressure end $(P \lesssim 10^{-6}\,\mathrm{mbar})$ and in the VF range $(P \gtrsim 10^{-1}\,\mathrm{mbar})$, the error increases, mainly due to a reduced sensitivity. 
Overall, $\Delta P/P<25\,\%$, except for the outliers at $P=\{3.9,\,9.5\}\,\mathrm{mbar}$ with $\Delta P/P=\{57,\,29\}\,\%$.

\subsection*{Conclusion}
In summary, we demonstrate a single nanomechanical gas pressure sensor, which exhibits, to the best of our knowledge, an unprecedented measurement range of 10 decades.
Combining an analytical model for molecular damping with FEM simulations (Comsol) for viscous damping, enables matching the measured pressure dependency of our sensor's mechanical quality factor within $15\,\%$, over the investigated pressure range.
This model thereby provides an accurate description of the transition between molecular and viscous flow regimes.
Furthermore, the FEM simulation reproduces the measured pressure dependency of the sensor's oscillation 
frequency in the viscous flow regime to within $4\,\%$. 
Our model relies only on known membrane parameters and gas type, together with values of the sensor's intrinsic resonance frequency and quality factor, which are measured at a pressure $<10^{-7}\,\mathrm{mbar}$. 
This self-calibration of the sensor significantly reduces the effort required compared to calibrating it via point-by-point comparison with a reference gauge \cite{ISO3567}.
The error of the pressure values inferred from quality factor and frequency measurements lies within $10\,\%$, for pressures between $\sim 10^{-6}\,\mathrm{mbar}$ and $\sim 10^{-1}\,\mathrm{mbar}$, and within $25\,\%$ for the complete measurement range from $10^{-7}\,\mathrm{mbar}$ to ambient pressure.
Exceptions are two data points at $P=\{3.9,\,9.5\}\,\mathrm{mbar}$ with related measurement errors of $\{57,\,29\}\,\%.$
We point out that the established accuracy, particularly of these outliers, might be limited by our commercial pressure gauge, which suggests comparing our sensor to a more accurate reference gauge, for determining its fundamental limitation.
Also, we discuss the non-ideal design of a membrane, as possible origin of deviations between data and model.
A possible route towards increased pressure sensitivity, and, possibly, reduced measurement error, appears to be the exploitation of the squeeze-film effect, resulting from installing an oscillating membrane at a $\mathrm{\mu}$m-scale distance from a neighboring surface \cite{salimi2023squeeze}.
Furthermore, the squeeze-film effect is expected to enable significantly (up to few orders of magnitude) increasing the damping acting on the membrane \cite{suijlen2009squeeze}.
This would translate to a corresponding reduction in the measurement time.
Increasing our sensor's intrinsic quality factor, e.g., by following a similar approach to Ref.~\cite{saarinen2022laser}, enables further extending its measurement range to ultra-high vacuum.
Furthermore, we expect our sensor to be suitable for measuring pressures significantly higher than ambient pressure, based on corresponding investigations for compact cantilevers~\cite{svitelskiy2009pressurized} (in our current setup, such investigations are prohibited by optical viewports incompatible with overpressure).
This, together with its inherent compatibility with cryogenic environments~\cite{zwickl2008high}, temperatures of several $100^{\circ}\mathrm{C}$~\cite{st2019swept}, other gases than air, magnetic fields~\cite{zwickl2008high}, and spatially constrained environments (when combined, e.g., with a compact readout via an optical fiber~\cite{flowers2012fiber}, on-chip photonics~\cite{guo2022integrated} or electronics~\cite{bagci2014optical}) illustrates the sensor's versatility. 

\section*{Acknowledgements}
The authors thank Vincent Dumont, Jack Sankey, Horst Schulte-Schrepping, and Detlef Sellman for helpful discussion, and Norcada Inc for providing great custom membrane solutions.

\section*{Author Information}
\subsection*{Author Contributions}
C.R. and H.M. contributed equally to this work.
C.R., H.M., J.S., A.L., and R.S. designed the research approach;
H.M. built the experiment with the help of C.R., and S.C;
H.M. performed the measurements with the help of C.R.;
C.R. developed the theoretical model;
H.M. C.R. and N.S. analyzed the data;
C.R., M.B.K.K., A.F., and H.M contributed to the simulations;
C.R., A.L, and R.S wrote the paper

\subsection*{Funding Sources}
This work was supported and partly financed by a PIER Seed Project PIF-2021-08 and by the DFG under Germany’s Excellence Strategy EXC 2121 ‘Quantum Universe’-390833306 (H.M., A.F.).

\bibliography{references}

\end{document}